# Aircraft Conflict Detection and Avoidance through Interacting Multiple Model (IMM) Estimation

Raja Manish and David Webster
*Graduate Students, Aeronautics and Astronautics Engineering*
*Purdue University, West Lafayette, USA*

Under
Inseok Hwang, *Ph.D.*
*Associate Professor, School of Aeronautics and Astronautics*
*Purdue University, West Lafayette, USA*

**The practical problem of tracking a maneuvering aircraft during flight has always been a crucial task in order to safeguard airborne assets from unknown threats. Therefore, the need for an efficient target detection and identification technique is substantial and growing. The multiple model (MM) estimation have proven to be one of the most reliable and accurate among various filtering algorithms. In this paper we will implement the Interacting Multiple Model (IMM) estimation technique for the aforementioned purpose of target identification. This target's motion, though defined by predefined dynamics, is obscured due to the noises from tracking sensors. The algorithm intends to predict the location of target and provide feedback maneuver to the reference aircraft in order to avoid a conflict.**

## Nomenclature

$\Pi$ = transition matrix
$r$ = number of modes
$\mu_i$ = mode probability
$\mu_{ij}$ = mixing probability
$x^{0j}$ = mixed Initial state
$P^{0j}$ = mixed initial covariance
$x^j(k|k)$ = mode-conditioned estimate of mode $j$ at time $k$
$P^j(k|k)$ = mode-conditioned covariance of mode $j$ at time $k$
$\Lambda_j(k)$ = likelihood function for mode $j$ at time $k$
$c_j$ = normalizing term
$v$ = process noise covariance
$w$ = measurement noise covariance
$S$ = variance of Gaussian pdf for likelihood function
$P$ = probability (0-1)

## I. Introduction

In this paper we present an IMM model based on a Jump Markov system for the purpose of identifying and avoiding aircraft conflicts. A three-mode Jump Markov system is presented and defined, followed by a definition of the probabilistic model of the state transitions. The IMM Model is then presented. Its benefits and downsides are discussed, as is the reasoning of choosing it over other multiple model estimators. The algorithm is then applied to an extension of the original problem, collision avoidance. The system is given a definition of what safe and unsafe flight of the tracked aircraft is, and a powerful avoidance algorithm is put into place, which uses rotation of the reference aircraft to avoid fatal crashes. The results of simulation are then presented and discussed.

## II. Jump Markov System Modeling

In order to establish the IMM estimator, we must model our system as a jump Markov system. Our system consists of the continuous state

$$X(k) \equiv [x_1(k), v_{x1}, x_2(k), v_{x2}, \Omega]^T$$

and the discrete state

$$q(k) = \{1, 2, 3\}$$

Where k in the time index for the discrete system. Our continuous state dynamics are given by:

$$X(k+1) = A_{q(k)} X(k) + B_{q(k)}$$

$$A_1 = \begin{bmatrix} 1 & \Delta T & 0 & 0 & 0 \\ 0 & 1 & 0 & 0 & 0 \\ 0 & 0 & 1 & \Delta T & 0 \\ 0 & 0 & 0 & 1 & 0 \\ 0 & 0 & 0 & 0 & 1 \end{bmatrix}, \quad A_2 = \begin{bmatrix} 1 & \frac{\sin(\beta + \pi/4)\Delta T}{\beta + \pi/4} & 0 & \frac{-(1 - \cos(\beta + \pi/4)\Delta T)}{\beta + \pi/4} & 0 \\ 0 & \cos(\beta + \pi/4)\Delta T & 0 & -\sin(\beta + \pi/4)\Delta T & 0 \\ 0 & \frac{-(1 - \cos(\beta + \pi/4)\Delta T)}{\beta + \pi/4} & 1 & \frac{\sin(\beta + \pi/4)\Delta T}{\beta + \pi/4} & 0 \\ 0 & \sin(\beta + \pi/4)\Delta T & 0 & \cos(\beta + \pi/4)\Delta T & 0 \\ 0 & 0 & 0 & 0 & 1 \end{bmatrix},$$

$$A_3 = \begin{bmatrix} 1 & \frac{\sin(\beta - \pi/4)\Delta T}{\beta - \pi/4} & 0 & \frac{-(1 - \cos(\beta - \pi/4)\Delta T)}{\beta - \pi/4} & 0 \\ 0 & \cos(\beta - \pi/4)\Delta T & 0 & -\sin(\beta - \pi/4)\Delta T & 0 \\ 0 & \frac{-(1 - \cos(\beta - \pi/4)\Delta T)}{\beta - \pi/4} & 1 & \frac{\sin(\beta - \pi/4)\Delta T}{\beta - \pi/4} & 0 \\ 0 & \sin(\beta - \pi/4)\Delta T & 0 & \cos(\beta - \pi/4)\Delta T & 0 \\ 0 & 0 & 0 & 0 & 1 \end{bmatrix}$$

$$B_1 = B_2 = B_3 = \begin{bmatrix} 0.5\Delta T^2 & 0 & 0 \\ \Delta T & 0 & 0 \\ 0 & 0.5\Delta T^2 & 0 \\ 0 & \Delta T & 0 \\ 0 & 0 & 0 \end{bmatrix}$$

The initial state $X_0$ is determined uniquely for each simulation. $[X_{1,0}, X_{2,0}]$ is a randomly determined point along a 4.5 km circle centered on the reference aircraft. $[Vx_{1,0}, Vx_{2,0}]$ is also randomized, but takes initial position into account, ensuring that the aircraft will quickly become unsafe if it continues straight. $\Omega_0$ is initially set to 0. Mode 1 is defined as straight, level flight according to the direction of the velocity vector. Mode 2 is a left turn and Mode 3 is a right turn.

When in either turn mode, the velocity vectors are changed for each time step spent turning. Then, if the aircraft were to transition back into straight flight, it would continue straight from the most recent incremental turn. This behavior is both more realistic, and more likely to produce accurate results than treating turns as static, instant changes. The $B_{q(k)}$ matrices are then used to include the noise and uncertainty values present in any observer-based system.

In the figure below, a simple overview of the coordinate system and system is presented, along with a visual illustration of the safe and unsafe regions of our state space.

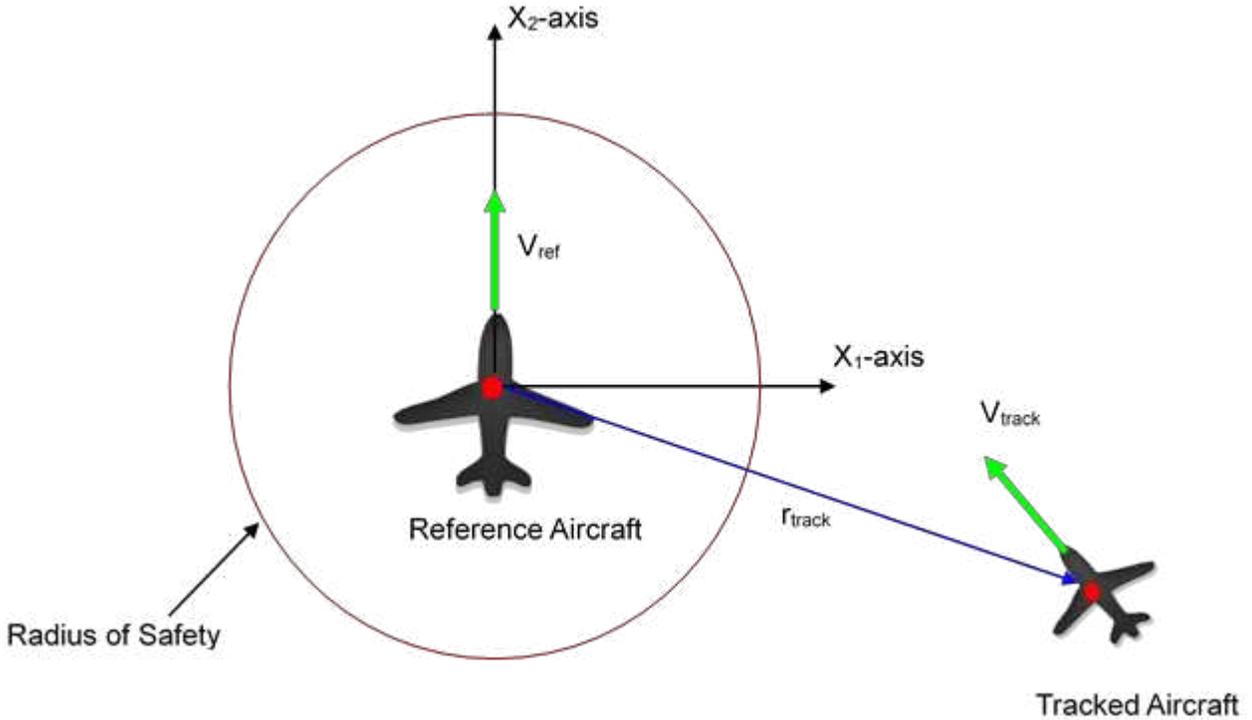

**Figure 1: Overview of coordinate system**

With the dynamics established, we define how the discrete state is governed. Discrete state transitions are determined by a finite state Markov chain with state space equivalent to Q.

The state probability vector is defined as

$$m(k) = \begin{bmatrix} m^1(k) & m^2(k) & m^3(k) \end{bmatrix}^T$$

Where $m^i(k), i = 1, 2, 3$, denotes the system is in the $i^{th}$ state at time k.

The discrete state transitions are governed by the evolution of the discrete state probability vector, given by:

$$m(k+1) = \Pi m(k)$$

Where $\Pi = [\pi_{ij}]_{3\times 3}$ is the transition probability matrix, with elements given by:
$$\pi_{ij} \equiv \Pr(q(k+1) = j | q(k) = i), \text{ with } \sum_{i=1}^{3} \pi_{ij} = 1 \text{ for } i = 1, 2, 3$$

For our particular system, we define $\Pi$ to be:

$$\Pi = \begin{bmatrix} .8 & .1 & .1 \\ .19 & .8 & .01 \\ .19 & .01 & .8 \end{bmatrix}$$

These values represent some fairly straightforward concepts. An aircraft is much more likely to stay in its current path than change. Therefore, for each mode, the probability that it will stay in that given mode in the next time step is much higher than anything else. For mode 1, it has an equal likelihood of making a left or right turn. For modes 2 and 3, the aircraft is more likely to return to traveling straight rather than going directly into, respectively, a right and left turn.

With continuous dynamics and evolution of the state probability vector defined, we now define the system as a whole, as shown in the figure below:

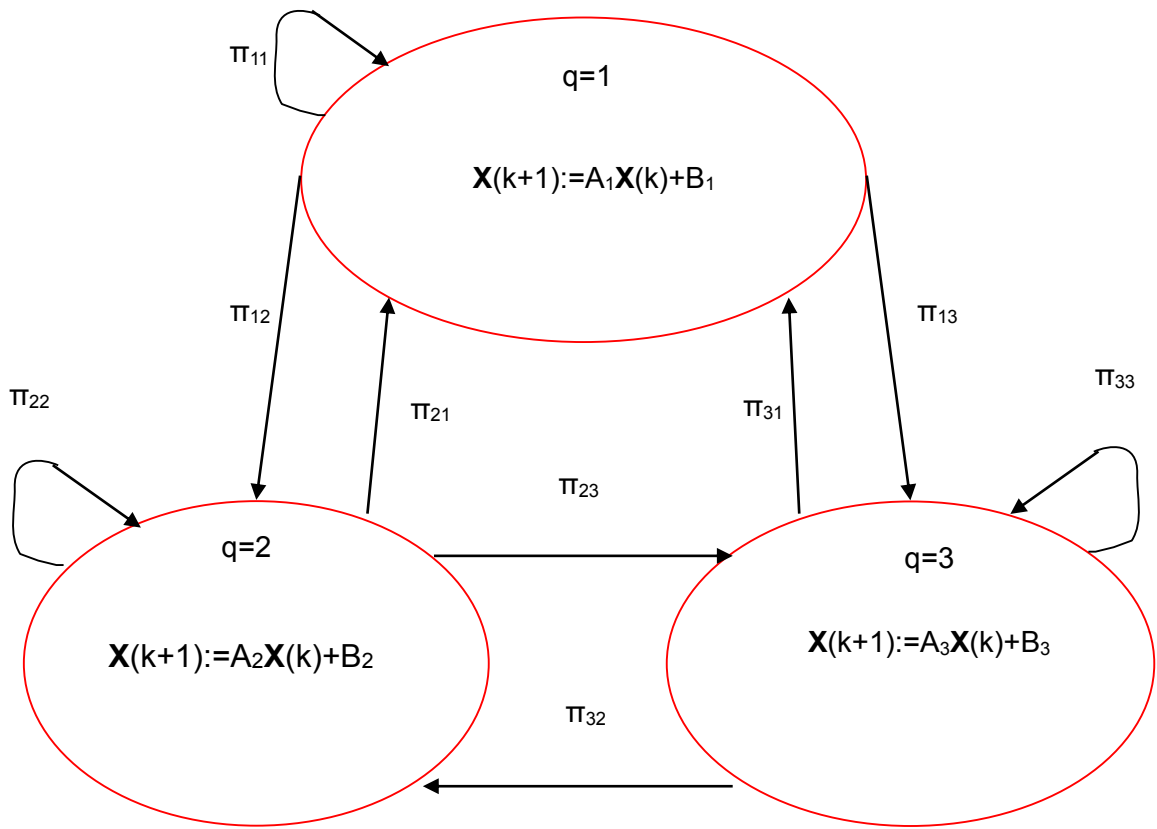

**Figure 2. Overview of System Dynamics**

## III. Background of Hybrid Estimation

The existence of various hybrid estimation techniques helps us in best determining the state of the system out of noisy measurements. Some of these have been developed in such a sophisticated manner that even with a random and intermittent availability of the sensor measurements, the estimation algorithm can give out the expected values of the state to be determined with high accuracy.

Few of the major estimation algorithms quite popular in industries are:

1. First-order Generalized Pseudo-Bayesian (GPB-I) approach
2. Second-order Generalized Pseudo-Bayesian (GPB-II) approach
3. Interacting Multiple Modelling (IMM) approach

The GPB I method considers each possible current model for the estimation of state. So it takes in calculation a total of $r$ possibilities.

In GPB II method, the state is computed under each possible current and previous models. Hence there are exactly $r^2$ possibilities which is taken in consideration.

The IMM algorithm computes the state estimate under each possible current model using $r$ filters with each filter using a different combination of the previous model conditioned estimates called as the mixed initial condition.

|  | GPB I | GPB II | IMM |
|---|---|---|---|
| Number of filters | $r$ | $r^2$ | $r$ |
| Number of combination of r estimates and covariance | 1 | $r+1$ | $r+1$ |
| Number of probability calculation | $r$ | $r^2+1$ | $r^2+1$ |

**Table 1: Comparison between the three estimation algorithms from *Bar-Shalom-Li*[1]**

Clearly, the IMM estimation technique is the best compromise between the performance of GPB I and the complexity of GPB II. That is, while maintaining the amount of accuracy similar to GPB II, IMM is able to able to reduce the computation expense of GPB II by performing mode-conditioned estimation.

Hence IMM was a clear choice for estimation in this paper.

**A. Need for IMM estimation**

A question arises that what is the need for IMM estimation and why can't we simply use a single mode Kalman filter. Firstly, when the uncertainty associated with the target is very high due to lack of the knowledge of its process, the estimation algorithm has to have high adaptive capability so as to capture as much information as possible.

The maneuverability is a function of uncertainty and the measurement. Higher the maneuverability, higher would be the uncertainty of the target and poorer would be the measurement by the same set of sensors. Hence an adaptive estimation technique such as IMM becomes important in this regard. It has been shown that above certain threshold of maneuverability, the performance of IMM is significantly better than Kalman filter given the measurements are appropriately interpreted.

Another reason could be that, a system with multiple discrete modes may have different process noises associated with each mode. This is true for example, an aircraft might have lower process noise when moving on linearly straight path while the noise may become large when taking a coordinated turn. A single model KF cannot handle multiple process noise declaration.

**B. Requirements for using IMM**
1. High maneuverability of the target – or the ratio of uncertainty of its position to the measurement.
2. Different process noise variance for different models – Also, the ratio between them should be large.
3. More than one mode should have non-zero likelihood – this is the outcome of different variances of the model.

For all the above three requirements, idea is that the algorithm demands more than one mode to have a certain likelihood of occurrence at a certain time in order for the algorithm to find a final estimate by considering weights of each mode. If this condition is not fulfilled, the IMM would become a disguised KF (page 475)[1].

## IV. State and Mode Estimation using IMM

The jump Markov model designed previously is now used with the Interacting Multiple Modelling (IMM) algorithm in order to determine the estimate of the position of target aircraft. The mode probability of the estimate indicates the type of movement our target aircraft is making. Since this system has 3 discrete modes, the intended IMM algorithm involves three Kalman filters which uses mixed initial estimate for estimating three new mode-conditioned estimate along with their probabilities. The following explains the IMM-KF estimation theory in brief.

**Interacting Multiple Model**

It is one of the most cost effective adaptive estimation technique for systems involving structural as well as parametric changes. It runs a bank of filters in parallel, each one based on a model matching to a particular mode of the system. In this project we are using 3 parallel Kalman filters for 3 modes. The overall state estimate is calculated by the probabilistically weighted sum of the outputs of all the 3 KFs. The interacting behavior is developed by explicitly modeling the abrupt changes of the system by "switching" from one model to another in a probabilistic manner.

The initial estimate at the beginning of each cycle for each filter is a mixture of all most recent estimates from the (iterative) single model based filters.

*Transition Probability:* - The change of state of the system are called transitions and the probabilities associated with various state changes are called Transition Probabilities. This process is characterized by a state space called Transition matrix.

*Markov chain*: - It is stochastic process with the Markov property. Markov property refers to the sequence of random variables moving through a process so that there is serial dependence only between adjacent period (what happens next depends only on current state of the system.)

The IMM algorithm used in this project has five states, two observed states and 3 filters each corresponding to each mode. The following four steps are the set of operations in each iteration:

1. *Mixing probability calculation*: It gives the probability that mode $M_i$ was active at time $i - 1$ given that mode $M_j$ is active at time $i$ conditioned on the measurement $Z_{i-1}$.
   Mathematically,
   $$\mu_{ij}(k-1|k-1) = \frac{\pi_{ij}\mu_i}{\sum_{i=1}^{r}\pi_{ij}\mu_i(k-1)}$$
   Where $\pi_{ij}$ = transition probability
   $\mu_i$ = mode probability

2. *Mixed Initial condition calculation*: Mixed initial condition for the filter matched to $M_j$ is computed using the previous mode estimates and mixing probabilities calculated above.
   $$x^{0j}(k-1|k-1) = \sum_{i=1}^{r} x^i(k-1|k-1)\mu_{ij}(k-1|k-1)$$
   Where $j = 1, ... r$
   $i = 1, ... r$

The corresponding covariance of above would be

$$P^{0j}(k-1|k-1) = \sum_{i=1}^{r} \left[ \begin{array}{c} \mu_{ij}(k-1|k-1)\{P^i(k-1|k-1) \\ + \\ [x^i(k-1|k-1) - x^{0j}(k-1|k-1)].[x^i(k-1|k-1) - x^{0j}(k-1|k-1)]^T\} \end{array} \right]$$

3. *Mode-matched filtering*: The mixed initial estimates and their corresponding covariances are now fed into the Kalman filter (KF) matched to mode $M_j(k)$ which uses the latest available measurement $z(k)$ to give $x^j(k|k)$ and $P^j(k|k)$. It also gives the likelihood $\Lambda_j(k)$ associated with each of the mode. This likelihood is a Gaussian pdf.

$$\Lambda_j(k) = \mathcal{N}[z(k); z^j[k|k-1; x^{0j}(k-1|k-1)], S^j[k; P^{0j}(k-1|k-1)]]$$

Where $S$ is the variance of the Gaussian pdf.

4. *Mode probability update*: Now the final probability that mode $M_j(k)$ is in effect is calculated using the likelihood computed in previous step conditioned on $Z^k$. It is given by

$$\mu_j(k) = \frac{\Lambda_j(k)\,\bar{c}_j}{\sum_{j=1}^{r} \Lambda_j(k)\,\bar{c}_j}$$

Where $\bar{c}_j = \sum_{i=1}^{r} \pi_{ij}\mu_i(k-1)$ as found while calculating mixing probabilities

*State Estimate and covariance computation:* The mode-conditioned estimates and covariances are combined using the mode probabilities and estimates and covariances from each KF by the following:

$$x(k|k) = \sum_{j=1}^{r} x^j(k|k)\mu_j(k)$$

$$P(k|k) = \sum_{j=1}^{r} \left[ \mu_j(k) \{ P^j(k|k) + [x^j(k|k) - x(k|k)].[x^j(k|k) - x(k|k)]^T \} \right]$$

Where $x^j(k|k)$ and $P^j(k|k)$ were obtained from each of the three KFs.

## V. Conflict Detection and Avoidance

With the IMM estimator in place, we have a powerful tool that, if employed properly, can provide us with a good prediction of where the tracked aircraft will be several time steps in the future. For our scenario, this is a very important extension of the IMM functionality because of the safety-critical nature of ATC.

We are assuming that the tracked aircraft is uncontrollable. This is a reasonable assumption based on the problem statement. The tracked aircraft has been given randomly determined flight paths and initial conditions, and has the real-world analogue of an aircraft with malfunctioning communication equipment or faulty navigation. It is infringing on the reference aircraft's airspace, with a very real threat of disastrous collision.

The conflict detection and avoidance (CDA) algorithm here will use the powerful prediction capabilities of the IMM to extrapolate the tracked aircraft's flight path forward and, if it lies within a predefined unsafe radius, alter the dynamics of the reference aircraft so that the predicted behavior of the tracked aircraft will become safe.

The algorithm looks up to three time-steps forward in time when searching for unsafe motion, but the error incurred in this process will still be on the scale of tens of meters, which is very small on the scale of kilometers. This error is further tolerable because of the constantly updating estimate and mode probabilities. The worst possible scenario would be an aircraft flying close to the unsafe radius, but not predicted to enter it. An aircraft in this situation could be deemed safe, but change modes suddenly and then be within the unsafe radius by the time the estimator detects it. However, the algorithm will immediately perform an escape maneuver on the next time step, continuing to execute maneuvers every time step until the aircraft's motion is considered safe.

This escape maneuver is accomplished by rotating the coordinate frame, since the reference aircraft is controllable, but also defines the coordinate frame. Once the rotation happens, the governing dynamics of the system must be redefined to account for this change. This happens immediately after the angle is determined. The prediction algorithm is shown in the following equations:

$$\Delta X_1 = X_1(k+1) - X_1(k)$$
$$\Delta X_2 = X_2(k+1) - X_2(k)$$

Then,

$$r_{pred} = \sqrt{\{[X_1(k) + j\Delta X_1]^2 + [X_2(k) + j\Delta X_2]^2\}},$$
$$\text{Where, } j = 1, 2, 3$$

Where $X_1$ and $X_2$ are the estimated state obtained from the IMM algorithm, and $r_{pred}$ is the distance from the origin, defined as the center of the aircraft body. The algorithm starts at $j = 1$, calculates where the aircraft is predicted to be, then compares this with the minimum safe distance, $r_{safe}$. If this distance does not violate any safety restrictions, the algorithm proceeds to $j = 2$, repeating this process up to $j = 3$. If at any point the aircraft is predicted to be within $r_{safe}$, the algorithm produces a warning and calculates an escape maneuver. The figure below demonstrates how the avoidance algorithm works.

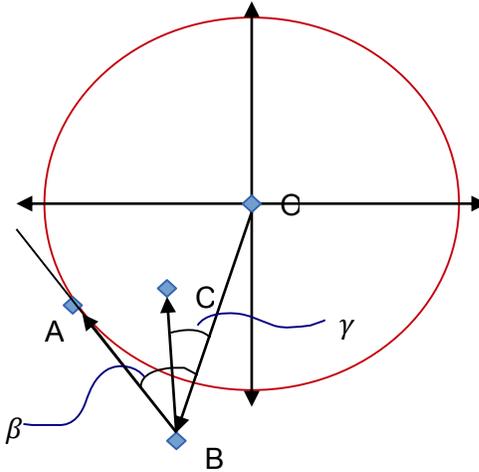

**Figure 3: Illustration of Avoidance Maneuver Calculation**

As can be seen in Figure 3, Point B is the current estimated location of the tracked aircraft. Point $C_j$ is the unsafe prediction of where it will be in $j$ time steps. Point A is a point on the Radius of Safety which, when connected with Point B, forms a line tangent to the surface of the Radius of Safety. $\beta$ is the angle formed between segments BA and BO. $\gamma$ is the angle formed between segments $BC_j$ and BO. Our desired angle is $\theta$. The algorithm will first seek to find θ such that, upon rotating the coordinate frame by the angle $\theta$, the unsafe prediction $C_j$ becomes tangent to the radius of safety. The algorithm will always find a real-valued answer for θ, however it will bound the answer to lie within the interval $[-\pi/4, \pi/4]$. This is done to avoid any unsafe or unrealistic suggested turn angles for our reference aircraft to perform in a single time step. The below equations will define all relevant calculations and quantities:

$$BO = norm_2(\boldsymbol{X}_{est}([1,3], i))$$
$$BC_j = norm_2\left(\boldsymbol{X}_{est}([1,3], i) + j\Delta T(\boldsymbol{X}_{est}([1,3], i+1) - \boldsymbol{X}_{est}([1,3], i))\right)$$
$$BA = BC_j$$
$$AO = r_{safe}$$
$$\angle ABO = \beta = \sin^{-1}\frac{AO}{BO}$$
$$\angle C_j BO = \gamma = \tan^{-1}\left(\frac{BC_j \times BO}{BC_j . BO}\right)$$
$$\theta = \beta - \gamma \text{ While } -\frac{\pi}{4} \leq \theta \leq \frac{\pi}{4}$$

If $\theta$ is outside either of its bounds, the algorithm limits it to the nearest boundary. This is, in effect, using the maximum safe turning value. The algorithm then rotates the coordinate frame by $\theta$ and reorients the dynamics to the new reference frame.

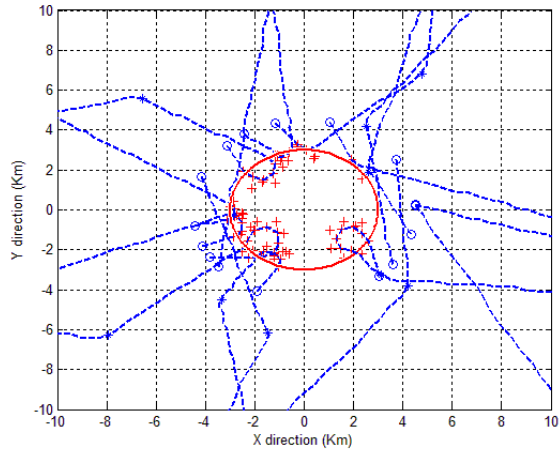

**Figure 4. Result of about 15 simulations**

## VI. Numerical simulation

The aircraft tracking algorithm presented here uses IMM to estimate the position coordinates and mode of the target with respect to the reference aircraft. The algorithm as described in the prediction and conflict avoidance section tries to maneuver the reference aircraft whenever the target is calculated to enter the unsafe perimeter. The algorithm is based on the following assumptions:

1. Trajectories of the target are generated in algorithm- it is random and does not follow any waypoints which could indicate the unpredictable behavior of, for example, a fighter aircraft.
2. Target does not change its mode while the reference aircraft is steered in the situation of a conflict.
3. The reference aircraft reorients while the target is still in mode 1, i.e. flying straight, although target may immediately switch its mode as soon as the reference aircraft finishes its maneuver over one time step.

4. When the target location is predicted, the probability that target would be at the predicted position keeps on decreasing as we predict more and more in future ($e.g. \Delta T \text{ to } 5\Delta T$). The algorithm does not takes into account this decreasing probability. Our prediction would only be up to $3\Delta T$.

Followings are the constants declared for the example:

$V_{cruise} = 285.841$ m/s, Nominal cruise speed for a jetliner, e.g. Boeing 747
$R_{safety} = 3000$ m, Radius of safety of reference aircraft

Please note that we have randomized the magnitude and direction of initial velocity so that the algorithm starts with different $\vec{V}$ for every trial execution, where
$$V \in [V_{cruise}, 2 \times V_{cruise}].$$

State estimate for the trajectory is generated with white noise using the process noise covariance matrix is given by

$$v = \begin{bmatrix} 200 & 0 & 0 & 0 & 0 \\ 0 & 0.1 & 0 & 0 & 0 \\ 0 & 0 & 200 & 0 & 0 \\ 0 & 0 & 0 & 0.1 & 0 \\ 0 & 0 & 0 & 0 & 0.001 \end{bmatrix}$$

Measurement of the coordinates is recorded with white noise using the covariance matrix given by,

$$w = \begin{bmatrix} 50^2 & 0 \\ 0 & 50^2 \end{bmatrix}$$

IMM uses a fixed Transition probability matrix $\Pi$ as defined initially.

*Process*: The target moves on its trajectory and tries to enter the safety perimeter of the reference. As soon as the algorithm predicts the possibility of the conflict, the correction algorithm tries to divert the reference aircraft from the target. After one time step, if the re-orientation maneuver is not sufficient to evade from the target, the algorithm tries to recalculate the new path for the reference to take.

## VII. Estimation result

Figure 5 shows the computed mode probabilities and the estimated mode by the algorithm. The change of mode was for one time step at a time. The mode probability plot as we can see in 5(a) is the outcome of the likelihoods of occurrence of each mode at any instant. The mode having highest probability is the mode in effect. We can see that the estimator has almost traced the actual mode of the track. One thing to notice is that the algorithm has detected mode 2 just before time, t = 10 seconds. By looking at figure 6 we can conclude that such a false detection could be because of the evading maneuver by the reference aircraft towards right (It should be kept in mind that figure 6 shows the relative track of the target; right maneuver of the reference to avoid conflict instead appears as target circling toward right.) Such a false detection could be filtered out by including the reference maneuver information in the estimator in order to cancel out its effect. False detection of mode 2 and mode 3 at other places could be attributed to the noise in the measurement and can be filtered out by setting a threshold of probability.

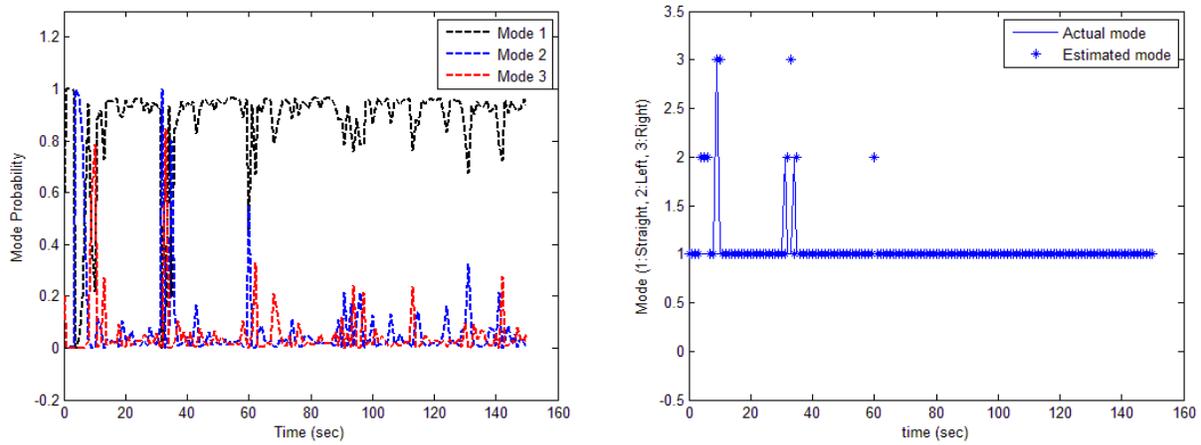

**(a)** Mode probability            **(b)** Estimated mode
**Figure 5. Mode probability and Mode estimation by the algorithm**

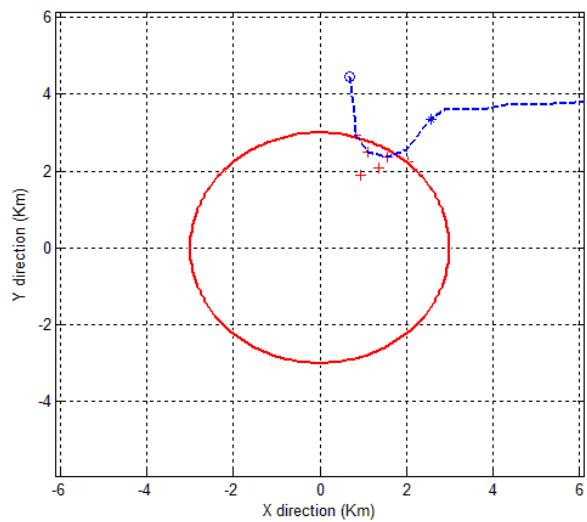

**Figure 6. Showing relative trajectory of the target in conflict and the radius of safety.**
Here, our target switches its mode to the right (indicated by blue '*') as soon as it is evaded. The red '+' sign on the track and corresponding '+' sign separate from the track, both as a pair (and other similar pairs) shows the predicted path of target along the same direction as its path in previous time step, identified by the algorithm to be inside the safety radius and hence unsafe for the reference aircraft.

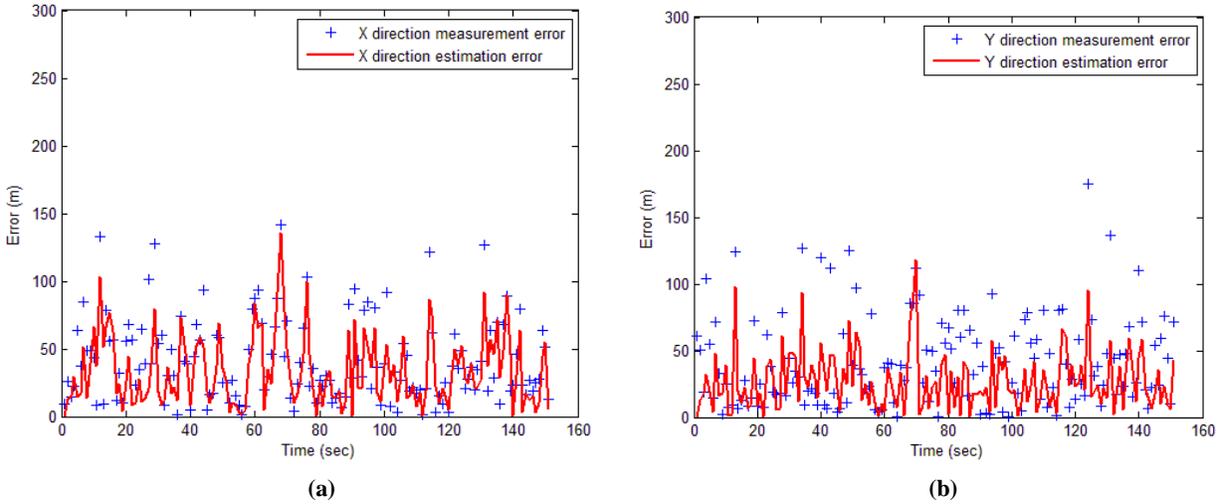

**Figure 7. Estimation error vs. measurement errors in (a) X direction and (b) Y direction**
**As expected, estimation error is lower and less spread than the measurement error.**

## VIII. Conclusion

A hybrid system with three switching modes was developed and represented with Markov blocks for its analysis. Using IMM estimation technique, we computed the estimates of the target aircraft with respect to the reference aircraft. Depending on the location of target, we predicted its position for next few time steps. This prediction was utilized in computing the reorientation value for the reference aircraft in order to avoid the conflict. The algorithm was tested to be working very well after numerous simulations. Having said so, one thing to note here is that the aircraft dynamic equations of motion used in the algorithm to track the target is based on the same dynamics with which the random target trajectory was generated. Although we have used standard flight dynamic equations in our algorithm and obtained very good results, it would be very interesting to find out the estimation efficiency of IMM along with the conflict avoidance when actual flight data from air traffic database are used instead of the one generated within the algorithm. In addition to this, the three discrete modes of flight in two dimensions could be improved to include all the three dimensions, i.e. in vertical plane. These two improvements could be one of our major work in the future.